\documentstyle[manuscript,aps]{revtex}
\begin{document}
\draft
\title{The averaged tensors of the relative energy-momentum and angular
momentum in general relativity}
\author{Janusz Garecki}
\address{Institute of Physics, University of Szczecin, Wielkopolska 15; 70-451
Szczecin, POLAND}
\date{\today}
\maketitle
\begin{abstract}
There exist at least a few different kind of averaging of the
differences of the energy-momentum and angular momentum in normal
coordinates {\bf NC(P)} which give tensorial quantities. The obtained
averaged quantities are equivalent mathematically because they differ
only by constant scalar dimensional factors.
One of these averaging was used in our papers [1-8] giving the {\it canonical
superenergy and angular supermomentum tensors}.

In this paper we present one
other averaging of the energy-momentum and angular momentum differences
which gives tensorial quantities with proper dimensions of the
energy-momentum and angular momentum densities. But these averaged
energy-momentum and angular momentum tensors, closely
related to the canonical superenergy and angular supermomentum
tensors, {\it depend on some fundamental length L}.

The averaged energy-momentum and angular momentum tensors of the
gravitational field obtained in the paper can be applied,
like the canonical superenergy and angular supermomentum tensors,
to coordinate independent local (and also global) analysis of this field.
\end {abstract}
\pacs{04.20.Me.0430.+x}

\section{The averaged energy-momentum and angular momentum
tensors in general relativity}
In the papers [1-8] we have defined the canonical superenergy and
angular supermomentum tensors, matter and gravitation, in general
relativity ({\bf GR}) and studied their properties and physical
applications. In the case of the gravitational field these tensors gave
us some substitutes of the non-existing gravitational
energy-momentum and gravitational angular momentum tensors.

The canonical superenergy and angular supermomentum tensors were
obtained pointwise as a result of some special averaging of the
differences of the energy-momentum and angular momentum in normal
coordinates {\bf NC(P)}. The dimensions of the components of these tensors can be written
down as:[the dimensions of the components of an energy-momentum or
angular momentum tensor (or pseudotensor)]$\times m^{-2}$.

In this paper we propose a new averaging of the energy-momentum and
angular momentum differences in {\bf NC(P)} which is very like to the
averaging used in [1-8] but which gives the averaged
quantities with proper dimensionality of the energy-momentum and angular
momentum densities.

Namely, we propose the following general definition of the averaged tensor
(or pseudotensor) density $\sqrt{\vert g\vert}T_a^b$
\begin{equation}
<{\it T_a^{~b}}(P)> := \displaystyle\lim_{\varepsilon\to
0}{\int\limits_{\Omega}{\bigl[{\it T_{(a)}^{~~~(b)}}(y) - {\it
T_{(a)}^{~~~(b)}}(P)\bigr]
d\Omega}\over\varepsilon^2/2\int\limits_{\Omega}d\Omega},
\end{equation}
where
\begin{equation}
{\it T_{(a)}^{~~~(b)}}(y) := (\sqrt{\vert g\vert}
T_i^{~k})(y){}e^i_{~(a)}(y){}e_k^{~(b)}(y),
\end{equation}
\begin{equation}
{\it T_{(a)}^{~~~(b)}}(P):= (\sqrt{\vert
g\vert}T_i^{~k})(P){}e^i_{~(a)}(P){}e_k^{~(b)}(P) = {\it T_a^{~b}}(P)
\end{equation}
are the tetrad (or physical) components of a tensor or a pseudotensor
density ${\sqrt{\vert g\vert} T_i^{~k}}(y)$ which describes an energy-momentum
distribution, $y$ is the collection of normal coordinates {\bf NC(P)} at
a given point {\bf P}, $e^i_{~(a)}(y),~e_k^{~(b)}(y)$ denote an
orthonormal tetrad field and its dual, respectively,
\begin{equation}
e^i_{~(a)}(P) = \delta^i_a,~e_k^{~(a)}(P)
=\delta^a_k,~e^i_{~(a)}(y)e_i^{~(b)}(y) =\delta_a^b,
\end{equation}
and they are parallelly propagated along geodesics through {\bf P}.

For a sufficiently small domain $\Omega$ which surrounds {\bf P} we
require
\begin{equation}
\int\limits_{\Omega}{y^id\Omega} = 0,~~\int\limits_{\Omega}{y^iy^kd\Omega} =
\delta^{ik} M,
\end{equation}
where
\begin{equation}
M = \int\limits_{\Omega}{(y^0)^2 d\Omega} = \int\limits_{\Omega} {(y^1)^2
d\Omega} = \int\limits_{\Omega}{(y^2)^2 d\Omega} =
\int\limits_{\Omega}{(y^3)^2 d\Omega},
\end{equation}
is a common value of the moments of inertia of the domain $\Omega$ with
respect to the subspaces $y^i = 0, ~i = 0,1,2,3$.

The procedure of averaging of an energy-momentum tensor or an
energy-momentum pseudotensor given in (1) is a four-dimensional
modification of the proposition by Mashhoon [9-12].

Let us choose $\Omega$ as a small analytic ball defined by
\begin{equation}
(y^0)^2 + (y^1)^2 + (y^2)^2 + (y^3)^2 \leq R^2 = \varepsilon^2 L^2,
\end{equation}
which can be described in a covariant way in terms of the auxiliary
positive-definite metric $h^{ik} := 2v^iv^k - g^{ik}$, where $v^i$ are
the components of the four-velocity of an observer {\bf O} at rest at
{\bf P} (see, e.g., [1-8]). $\varepsilon$ means a small parameter: $\varepsilon\in (0;1)$
and $L$ is a fundamental length.

Since at {\bf P} the tetrad and normal components are equal, from now on
we will write the components of any quantity at {\bf P} without (tetrad)
brackets, e.g., $T_a^{~b}(P)$ instead of $T_{(a)}^{~~~(b)}(P)$ and so on.

Let us now make the following expansions for the energy-momentum tensor
of matter $T_i^{~k}(y)$ and for $\sqrt{\vert g\vert}, e^i_{~(a)}(y),
e_k^{~(b)}(y)$ [13]
\begin{eqnarray}
T_i^{~k}(y) &=& {\hat T}_i^{~k} + \nabla_l {\hat T}_i^{~k}y^l +1/2{{\hat
T}_i^{~k}}{}_{,lm} y^ly^m + R_3\nonumber\\
&=& {\hat T}_i^{~k} +\nabla_l{\hat T}_i^{~k} y^l
+1/2\biggl[\nabla_{(l}\nabla_{m)}{\hat T}_i^{~k}\nonumber\\
&-&1/3{\hat R}^c_{~(l\vert i\vert m)}{}{\hat T}_c^{~k} +1/3 {\hat
R}^k_{~(l\vert c\vert m)}{} {\hat T}_i^{~c}\biggr] y^ly^m + R_3,
\end{eqnarray}
\begin{equation}
\sqrt{\vert g\vert} = 1 - 1/6 {\hat R}_{ab}y^ay^b + R_3
\end{equation}
\begin{equation}
e^i_{~(a)}(y) = {\hat e}^i_{~(a)} + 1/6{\hat R}^i_{~lkm}{\hat
e}^k_{~(a)} y^ly^m + R_3,
\end{equation}
\begin{equation}
e_k^{~(b)}(y) = {\hat e}_k^{~(b)} -1/6{\hat R}^p_{~lkm}{\hat e}_p^{~(b)}
y^ly^m + R_3,
\end{equation}
which give (1) in the form
\begin{equation}
<_m T_a^{~b}(P)> =\displaystyle\lim_{\varepsilon\to
0}{\int\limits_{\Omega}\bigl[\nabla_l{\hat T}_a^{~b} y^l +
1/2\bigl(\nabla_{(l}\nabla_{m)}{\hat T}_a^{~b} -2/3{\hat R}_{lm}{\hat
T}_a^{~b}\bigr)y^ly^m + THO\bigr]d\Omega\over
\varepsilon^2/2\int\limits_{\Omega}d\Omega}
\end{equation}
where $THO$ means the terms of higher order in the expansion of the
differences $T_{(a)}^{~~~(b)}(y) - T_{(a)}^{~~~(b)}(P) =
T_{(a)}^{~~~(b)}(y) - T_a^{~b}(P)$, $R_3$ is the remainder of the third
order and  $\nabla$ denotes covariant differentiation. Hat denotes
the value of an object at {\bf P}.

The first and $THO$ terms in the numerator of (12) do not contribute to
$<_m T_a^{~b}(P)>$. Hence, we finally get from (12)
\begin{equation}
<_m T_a^{~b}(P)> = _m S_a^{~b}(P) {L^2\over 6},
\end{equation}
where
\begin{equation}
_m S_a^{~b}(P) := \delta^{mn}\bigl[\nabla_{(l}\nabla_{m)}{\hat T}_a^{~b}
-2/3 {\hat R}_{lm}{\hat T}_a^{~b}\bigr]
\end{equation}
is the {\it canonical superenergy tensor of matter}\footnote{This tensor
is something different than the tensor given in [1-8] because here we
have averaged the tensor density $\sqrt{\vert g\vert}T_i^{~k}$; not the
tensor $T_i^{~k}$ as in [1-8]} [1-8].

By introducing the four velocity ${\hat v}^l \dot = \delta^l_0,~v^lv_l =1$ of
an observer {\bf O} at rest at {\bf P} and the local metric ${\hat
g}^{ab}\dot = \eta^{ab}$, where $\eta^{ab}$ is the inverse Minkowski
metric, one can write (14) in a covariant way as
\begin{equation}
_m S_a^{~b} (P;v^l) = \bigl(2{\hat v}^l{\hat v}^m -
{\hat g}^{lm}\bigr)\bigl[\nabla_{(l}\nabla_{m)} {\hat T}_a^{~b} -2/3{\hat
R}_{lm}{\hat T}_a^{~b}\bigr].
\end{equation}
The sign $\dot =$ means that an equality is valid only in some special
coordinates.

The matter superenergy tensor $_m S_a^{~b}(P;v^l)$ {\it is symmetric}.
As a result of an averaging the tensor $_m S_a^{~b}(P;v^l)$, and in
consequence the averaged tensor $<_m T_a^{~b}(P;v^l)>$, do not satisfy any
local conservation laws in general relativity. However, these tensors satisfy
trivial local conservation laws\footnote{Trivial local conservation laws because the integral
superenergetic quantities or, equivalently, integral averaged energy-momentum calculated from
them for aclosed system in special relativity vanish.} in special relativity (see,
e.g., [1-8]).

Now let us take the gravitational field and make the expansion
\begin{eqnarray}
\sqrt{\vert g\vert}_E t_i^{~k}(y) &=& {\alpha\over 9}\biggl[{\hat
B}^i_{~ilm} + {\hat P}^k_{~ilm}\nonumber\\
&-& {\delta_i^k\over 2}{\hat R}^{abc}_{~~~l}\bigl({\hat R}_{abcm} +
{\hat R}_{acbm}\bigr) + 2\beta^2\delta_i^k{\hat E}_{(l\vert g} {\hat
E}^g_{~\vert m)}\nonumber\\
&-& 3\beta^2{\hat E}_{i(l\vert}{\hat E}^k_{~\vert m)} + 2\beta{\hat
R}^k_{~(gi)(l\vert}{\hat E}^g_{~\vert m)}\biggr] y^ly^m + R_3.
\end{eqnarray}
Here
\begin{equation}
\alpha = {c^4\over 16\pi G} = {1\over 2\beta},~E_i^{~k} := T_i^{~k} -
1/2\delta_i^k T.
\end{equation}

The expansion (16) with the help of (9),(10) and (11) gives the following
averaged gravitational energy-momentum tensor
\begin{equation}
<_g t_a^{~b}(P;v^l)> = _g S_a^{~b}(P;v^l){L^2\over 6},
\end{equation}
where the tensor $_g S_a^{~b}(P;v^l)$ is the {\it canonical superenergy
tensor} for the gravitational field [1-8].

We have [1-8]
\begin{eqnarray}
_g S_a^{~b}(P;v^l) &=& {2\alpha\over 9}\bigl(2{\hat v}^l{\hat v}^m -
{\hat g}^{lm}\bigr)\biggl[{\hat B}^b_{~alm} + {\hat
P}^b_{~alm}\nonumber\\
&-& 1/2\delta_a^b{\hat R}^{ijk}_{~~~m}\bigl({\hat R}_{ijkl} + {\hat
R}_{ikjl}\bigr) + 2\beta^2\delta_a^b{\hat E}_{(l\vert g}{\hat
E}^g_{~\vert m)}\nonumber\\
&-& 3\beta^2{\hat E}_{a(l\vert}{\hat E}^b_{~\vert m)} + 2\beta{\hat
R}^b_{~(ag)(l\vert} {\hat E}^g_{~\vert m)}\biggr],
\end{eqnarray}
where
\begin{equation}
B^b_{~alm} := 2 R^{bik}_{~~~(l\vert} R_{aik\vert m)} -1/2\delta_a^b
R^{ijk}_{~~~l}{} R_{ijkm},
\end{equation}
is the {\it Bel-Robinson} tensor, while
\begin{equation}
P^b_{~alm} := 2 R^{bik}_{~~~(l\vert} R_{aki\vert m)} -1/2\delta_a^b
R^{ijk}_{~~~~l}{} R_{ikjm}.
\end{equation}

In vacuum the tensor $_g S_a^{~b}(P;v^l)$ reduces to the simpler form
\begin{equation}
_g S_a^{~b} (P;v^l) = {8\alpha\over 9}\bigl(2{\hat v}^l{\hat v}^m -
{\hat g}^{lm}\bigr)\biggl[{\hat R}^{b(ik)}_{~~~~~(l\vert}{\hat
R}_{aik\vert m)} -1/2\delta_a^b{\hat R}^{i(kp)}_{~~~~~(l\vert}{\hat
R}_{ikp\vert m)}\biggr],
\end{equation}
which is symmetric and the quadratic form $_g S_{ab}(P;v^l){\hat
v}^a{\hat v}^b$ is {\it positive-definite}.

In vacuum we also  have the {\it local conservation laws}
\begin{equation}
\nabla_b~{_g {\hat S}_a^{~b}} = 0.
\end{equation}
and the analogical laws satisfied by  the averaged tensor $<_g
t_a^{~b}(P;v^l)>$.

The averaged energy-momentum tensors $<_m T_a^{~b}(P;v^l)>$ and $<_g
t_a^{~b}(P;v^l)>$ can be considered as the {\it averaged tensors of the relative
energy-momentum}. They can also be interpreted as the {\it fluxes} of the
appropriate canonical superenergy. It is easily seen from the
formulas (13) and (18).

Now let us consider the {\it averaged angular momentum tensors} in {\bf
GR}.  The constructive definition of these tensors, in analogy to the
definition of the averaged energy-momentum tensors, is as follows.

In normal coordinates {\bf NC(P)} we define
\begin{equation}
<M^{(a)(b)(c)}(P)> = <M^{abc}(P)> :=\displaystyle\lim_{\varepsilon\to
0}{ \int\limits_{\Omega}{\bigl[M^{(a)(b)(c)}(y) -
M^{(a)(b)(c)}(P)\bigr]d\Omega}\over\varepsilon^2/2\int\limits_{\Omega}d\Omega},
\end{equation}
where
\begin{equation}
M^{(a)(b)(c)}(y) :=
M^{ikl}(y){}e_i^{~(a)}(y){}e_k^{~(b)}(y){}e_l^{~(c)}(y),
\end{equation}
\begin{equation}
M^{(a)(b)(c)}(P) := M^{ikl}(P)
e_i^{~(a)}(P){}e_k^{~(b)}(P){}e_l^{~(c)}(P) =
M^{ikl}(P)\delta_i^a\delta_k^b\delta_l^c = M^{abc}(P),
\end{equation}
are the {\it physical} (or tetrad) components of the field $M^{ikl}(y)=
(-)M^{kil}(y)$ which describes the angular momentum densities
\footnote{ Of course, $M^{abc}(P) = 0$, but we leave $M^{abc}(P)$ in our
formulas.}.
As in (2) and (3) , $e^i_{~(a)}(y),~~e_k^{~(b)}(y)$ denote mutually dual
orthonormal tetrads parallelly propagated along geodesics through {\bf
P} such that $e^i_{~(a)}(P) = \delta^i_a,~~e_k^{~(b)}(P) = \delta_k^b$.
$\Omega$ is a sufficiently small four-dimensional ball with centre at
{\bf P} and with the radius $R = \varepsilon L$.

At {\bf P} the tetrad and normal components of an object are equal. We
apply this again and omit tetrad brackets for the indices of any
quantity attached to the point {\bf P}; for example, we write
$M^{abc}(P)$ instead of $M^{(a)(b)(c)}(P)$ and so on.

For matter as $M^{ikl}(y)$ we take
\begin{equation}
_m M^{ikl}(y) =\sqrt{\vert g\vert}\bigl[y^i T^{kl}(y) - y^k
T^{il}(y)\bigr],
\end{equation}
where $T^{ik}(y) = T^{kl}(y)$ are the components of a symmetric
energy-momentum tensor of matter and $y^i$ denote the normal coordinates
{\bf NC(P)}.

The formula (27) gives the total angular momentum densities, orbital and
spinorial, because the dynamical energy-momentum tensor of matter
$T^{ik} = T^{ki}$ comes from the canonical one by
using the {\it Belinfante-Rosenfeld} symmetrization procedure and,
therefore, includes the canonical spin of matter [14].

Note that the normal coordinates $y^i$ form the components of the local
radius-vector ${\vec y}$ with respect to the origin {\bf P}.
Consequently, the components $_m M^{ikl}(y)$ form a local tensor density.

For the gravitational field we take the gravitational angular momentum
pseudotensor proposed by Bergmann and Thomson [14,17] as
\begin{equation}
_g M^{ikl}(y) = _F U^{i[kl]}(y) - _F U^{k[il]}(y)+\sqrt{\vert
g\vert}\bigl(y^i _{BT} t^{kl} - y^k _{BT} t^{il}\bigr),
\end{equation}
where
\begin{equation}
_F U^{i[kl]}:= g^{im}{}_F U_m^{~[kl]} =\alpha
g^{im}{g_{ma}\over\sqrt{\vert g\vert}}\biggl[(-g)\bigl(g^{ka} g^{lb} -
g^{la} g^{kb}\bigr)\biggr]_{,b}
\end{equation}
are {\it Freud's superpotentials} with the first index raised and
\begin{equation}
_{BT} t^{kl}:= g^{ki}{} _Et_i^{~l} + {g^{mk}_{~~,p}\over\sqrt{\vert
g\vert}} {} _F U_m^{~[lp]}
\end{equation}
are the components of the {\it Bergmann-Thomson} gravitational
energy-momentum pseudotensor [14,17].
\begin{eqnarray}
_E t_i^{~k} &=& \alpha\biggl\{\delta_i^k
g^{ms}\bigl(\Gamma^l_{~mr}\Gamma^r_{~sl} - \Gamma^r_{~ms}
\Gamma^l_{~rl}\bigr)\nonumber\\
&+& g^{ms}_{~~,i}\bigl[\Gamma^k_{~ms} - 1/2\bigl(\Gamma^k_{~tp}{}g^{tp}
- \Gamma^l_{~tl}{} g^{kt}\bigr)g_{ms}\nonumber\\
&-&1/2\bigl(\delta_s^k \Gamma^l_{~ml} + \delta^k_m
\Gamma^l_{~sl}\bigr)\biggr]\biggr\}
\end{eqnarray}
is the {\it Einstein canonical gravitational energy-momentum
pseudotensor} of the gravitational field.

The Bergmann-Thomson gravitational angular pseudotensor is most closely
related to the Einstein canonical energy-momentum complex and it has
better physical and transformational properties than the famous
gravitational angular momentum pseudotensor proposed by Landau and
Lifschitz [15,16,17]. This is why we apply it here.

One can interpret the Bergmann-Thomson gravitational angular momentum
pseudotensor as the sum of the {\it spinorial part}
\begin{equation}
S^{ikl} := _F U^{i[kl]} - _F U^{k[il]}
\end{equation}
and the {\it orbital part}
\begin{equation}
O^{ikl} := \sqrt{\vert g\vert}\bigl(y^i{}_{BT} t^{kl} - y^k{}_{BT}
t^{il}\bigr)
\end{equation}
of the gravitational angular momentum ``densities''.

Substitution of (27) and (28) (expanded up to third order) and
(9),(10),(11) into (24) gives the following {\it averaged angular momentum
tensors} for matter and gravitation respectively
\begin{equation}
<_m M^{abc}(P;v^l)> = _m S^{abc}(P;v^l){L^2\over 6},
\end{equation}
\begin{equation}
<_g M^{abc}(P;v^l)> = _g S^{abc}(P;v^l){L^2\over 6}.
\end{equation}

Here
\begin{equation}
_m S^{abc}(P;v^l) = 2\bigl[\bigl(2{\hat v}^a{\hat v}^p - {\hat
g}^{ap}\bigr)\nabla_p{\hat T}^{bc} - \bigl(2{\hat v}^b{\hat v}^p - {\hat
g}^{bp}\bigr) \nabla_p {\hat T}^{ac}\bigr],
\end{equation}
and
\begin{eqnarray}
_g S^{abc}(P;v^l) &=& \alpha\bigl(2{\hat v}^p{\hat v}^t - {\hat
g}^{pt}\bigr)\biggl[\beta\bigl({\hat g}^{ac}{\hat g}^{br} -{\hat
g}^{bc}{\hat g}^{ar}\bigr)\nabla_{(t}{\hat E}_{pr)}\nonumber\\
&+& 2{\hat g}^{ar}\nabla_{(t}{\hat R}^{(b}_{~~p}{}^{c)}_{~~r)} - 2{\hat
g}^{br} \nabla_{(t}{\hat R}^{(a}_{~~p}{}^{c)}_{~~r)}\nonumber\\
&+& 2/3{\hat g}^{bc}\bigl(\nabla_r{\hat R}^r_{~(t}{}^{a}_{~p)}
-\beta\nabla_{(p} {\hat E}^a_{~t)}\bigr) -2/3{\hat g}^{ac}\bigl(\nabla_r
{\hat R}^r_{~(t}{}^b_{~p)} - \beta\nabla_{(p} {\hat
E}^b_{~t)}\bigr)\biggr]
\end{eqnarray}
are the components of the {\it canonical angular supermomentum tensors} for
matter and gravitation, respectively [4,6,8].

In special relativity the averaged tensors $<_g M^{abc}(P;v^l)>, ~<_m
M^{abc}(P;v^l)>$, like as the canonical angular supermomentum tensors,
satisfy trivial conservation laws [1-8]. In the framework of the {\bf GR}
only the tensors $_g S^{abc}(P;v^l)$ and $<_g M^{abc}(P;v^l)>$ satisfy
local conservation laws in vacuum.

In vacuum, when $T_{ik} = 0 \Longleftrightarrow E_{ik} := T_{ik} -1/2
g_{ik}T = 0$, the canonical gravitational angular supermomentum tensor
$_g S^{abc}(P;v^l) = (-) _g S^{bac}(P;v^l)$ given by (37) simplifies to
\begin{equation}
_g S^{abc}(P;v^l) =2\alpha\bigl(2{\hat v}^p{\hat v}^t - {\hat
g}^{pt}\bigr)\biggl[{\hat g}^{ar}\nabla_{(p}{\hat R}^{(b}_{~~t}{}^{c)}_{~~r)}
-{\hat g}^{br}\nabla_{(p} {\hat R}^{(a}_{~~t}{}^{c)}_{~~r)}\biggr].
\end{equation}

Some remarks are in order:
\begin{enumerate}
\item The orbital part $O^{ikl} =\sqrt{\vert g\vert}\bigl(y^i_{BT}
t^{kl} - y^k _{BT} t^{il}\bigr)$ of the $_g M^{ikl}$ {\it does not
contribute} to the tensor $_g S^{abc}(P;v^l)$ and to the tensor $<_g
M^{abc}(P;v^l)>$. Only the spinorial part $S^{ikl} = _F U^{i[kl]} - _F
U^{k[il]}$ gives nonzero contribution to these tensors.
\item The averaged angular nomentum tensors $<_gM^{abc}(P;v^l)>, ~~<_m
M^{abc}(P;v^l)>$,like as the canonical angular supermomentum tensors,
{\it do not need} any radius-vector for existing.
\end{enumerate}

The averaged tensors $< _m M^{abc}(P;v^l)>,~~<_g M^{abc}(P;v^l)>$,
likely as the averaged energy-momentum tensors, can be interpreted as
the {\it averaged tensors of the relative angular
momentum};\footnote{The angular momentum is, of course, always relative
quantity, by definition.} and also as the {\it fluxes} of the
appropriate angular supermomentum.

The formulas (13),(18),(34) and (35) give the direct link beteween
the canonical superenergy and angular supermomentum tensors
\begin{equation}
 _gS_a^{~b}(P;v^l),~_m S_a^{~b}(P;v^l), ~_g S^{abc}(P;v^l), ~~_m
S^{abc}(P;v^l)
\end{equation}
and the averaged energy-momentum and angular momentum tensors
\begin{equation}
<_g t_a^{~b}(P;v^l)>, <_m T_a^{~b}(P;v^l)>, <_g
M^{abc}(P;v^l)>, <_m S^{abc}(P;v^l)>.
\end{equation}

It is easily seen that the averaged energy-momentum and angular momentum
tensors {\it differ} from the canonical superenergy and angular supermomentum tensors
{\it only} by the constant scalar multiplicator ${L^2\over 6}$, where $L$ means
some fundamental length. Thus, from the mathematical point of view, these two kind
of tensors are equivalent. Physically they {\it are not} because their
components have different dimension. Moreover the averaged
energy-momentum and angular momentum tensors depend on a fundamental
length $L$. Owing to the last fact and the formulas (13),(18), (34), (35) it seems that
the canonical superenergy and angular supermomentum tensors are {\it more fundamental}
than the averaged energy-momentum and angular momentum tensors. But one should emphasize
that the averaged energy-momentum and angular momentum tensors have an important
superiority over the canonical superenergy and angular supermomentum
tensors: their components {\it possesse proper dimensions} of the
energy-momentum and angular momentum densities.

The averaged tensors
\begin{equation}
<_g t_a^{~b}(P;v^l)>, ~<_m T_a^{~b}(P;v^l)>,~~ <_g
M^{abc}(P;v^l)>, ~<_m M^{abc}(P;v^l)>
\end{equation}
depend on the four-velocity ${\vec v}$ of a fiducial observer {\bf O} which is at rest at
the beginning {\bf P} of the normal coordinates {\bf NC(P)} used for
averaging and on some fundamental length $L$. After fixing of the length
$L$ one can determine univocally these tensors along the world line of
an observer {\bf O}.

In general one can {\it unambiguously determine} these tensors (after fixing
$L$) in the whole spacetime or in some domain $\Omega$ if in the spacetime or
in the domain $\Omega$ a geometrically distinguised timelike unit vector
field ${\vec v}$ is given.

One can try to establish\footnote{But this is {\it not necessary}. One can effectively use
the averaged energy-momentum and angular momentum tensors {\it without fixing L}.}
the fundamental length $L$ by using loop quantum gravity.
Namely, one can take as $L$ e.g., the smallest length $l$ over which the
classical model of the spacetime is admissible.
\par
Following loop quantum gravity [18-28] one can say about continuous classical
differential geometry already just a few orders of magnitude above the Planck
scale, e.g., for distances $l\geq 100L_P = 100\sqrt{{G\hbar\over c^3}}
\approx 10^{-33}$ m. So, one can take as the fundamental length $L$ the
value $L = 100 L_P \approx 10^{-33}$ m.

After fixing the fundamental length $L$ we have the averaged
energy-momentum and angular momentum tensors established with the same precise
as the canonical superenergy and angular supermomentum tensors.
\par
The averaged tensors (with $L$ fixed or no)
\begin{equation}
<_mT_a^{~b}(P;v^l)>, <_g t_a^{~b}(P;v^l)>, <_m M^{abc}(P;v^l)>, <_g
M^{abc}(P;v^l)>
\end{equation}
give us as good tool to a local ( and also global) analysis
of the gravitational and matter fields as the canonical superenergy and
angular supermomentum tensors
\begin{equation}
_m S_a^{~b}(P;v^l), ~_g S_a^{~b}(P;v^l),
~_m M^{abc}(P;v^l), ~_g M^{abc}(P;v^l)
\end{equation}
give. For example, one can apply the averaged energy-momentum and
angular momentum tensors to the all problems which have been analyzed in the papers
[1-8].

In this paper we only apply the averaged gravitational energy-momentum tensor\hfill\break
$<_g t_a^{~b}(P;v^l)>$ in order to decide if free vacuum
gravitational field has any kind of energy-momentum; especially, if
gravitational waves carry any energy-momentum? The problem arose
recently because some authors conjectured [29-33], by using
coordinate dependent pseudotensors and complexes, that the energy and momentum in
general relativity are confined to the regions of non-vanishing
energy-momentum tensor of matter and that the gravitational waves carry no energy
and momentum.
The argumentation is the following. For some solutions to the Einstein
equations and in some special coordinates, e.g., in Bonnor's spacetime
[34] in Bonnor's or in Kerr-Schild coordinates, the Einstein canonical
gravitational energy-momentum pseudotensor (and other pseudotensors
also) {\it globally vanishes} outside of the domain in which
$T^{ik}\not= 0$. The analogical global vanishing of the canonical
pseudotensor $_E t_a^{~b}$ we have for the plane and for the
plane-fronted gravitational waves in, e.g., null coreper [3,35].
But one should emphasize that all these results are {\it coordinate
dependent} [3,7,35]. Moreover, one should interpret physically the global
vanishing of the canonical pseudotensor (and other pseudotensors also) in some coordinates in
vacuum as a {\it global cancellation} of the energy-momentum of the real gravitational
field which has $R_{iklm}\not= 0$ with energy-momentum of the inertial forces field
which has $R_{iklm}=0$; {\it not as a proof of vanishing of the energy-momentum  of the real
gravitational field}. It is because the all used pseudotensors were entirely constructed from
the Levi-Civita's connection $\Gamma^i_{~kl} = \Gamma^i_{~lk}$ which describes a mixture of the
real gravitational field ($R_{iklm}\not= 0$) and an inertial forces field ($R_{iklm}= 0$).

In order to get the coordinate independent results about
energy-momentum of the {\it the real gravitational field}  one must use
tensorial expressions which depend on curvature tensor, like the averaged
gravitational energy-momentum tensor $<_g t_a^{~b}(P;v^l)>$ or like the
canonical gravitational superenergy tensor $_g S_a^{~b}(P;v^l$). These two tensors
vanish iff $R_{iklm}=0$, i.e., iff the spacetime is flat and we have no real
gravitational field.

When calculated, the averaged gravitational energy-momentum tensor $<_g
t_a^{~b}(P;v^l)>$ always gives the {\it positive-definite} averaged free relative
gravitational energy density and, in the case of the gravitational waves, its non-zero flux.
It is easily seen from the our papers [1-8,35].

Thus, the conjecture about localization of the gravitational energy only
to the regions of the non-vanishing energy-momentum tensor of matter
{\it cannot be correct} for the real gravitational field which has
$R_{iklm}\not= 0$.

In a similar way one can use the averaged gravitational angular momentum
tensor \hfill\break $<_g M^{abc}(P;v^l)>$ to coordinate-independent analysis of
the angular momentum of the real gravitational field.

\end{document}